# Above-Barrier Reflection of Cold Atoms by Resonant Laser Light within the Gross–Pitaevskii Approximation[1]

H. A. Ishkhanyan and V. P. Krainov

*Moscow Institute of Physics and Technology, Dolgoprudny, Moscow Region, 141700 Russia*
e-mail: vpkrainov@mail.ru

**Abstract**—Above-barrier reflection of cold alkali atoms by resonant laser light was considered analytically within the Gross–Pitaevskii approximation. Correction for the reflection coefficient because of a weak nonlinearity of the stationary Schrödinger equation has been derived using multiscale analysis as a form of perturbation theory. The nonlinearity adds spatial harmonics to linear incident and reflecting waves. It was shown that the role of nonlinearity increases when the kinetic energy of an atom is nearly to the height of the potential barrier. Results are compared to the known numerical derivations for wave functions of the Gross–Pitaevskii equation with the step potential.

PACS numbers: 03.750.Fu, 05.30.Jp, 67.40.-w

**DOI:** 10.1134/S1054660X09160038

In this paper we consider the problem in which an atom (for example, Bose isotope of the alkali atom $^{87}$Rb with the nuclear spin of 3/2 in gaseous medium of the same atoms [1]) moves slowly oppositely to the focused laser beam. The laser frequency is supposed to be equal to the frequency of dipole transition of this atom to the first excited state (Fig. 1). In the focusing region resonant absorption of laser photons occurs. The absorption probability can be equal to 100% under some special choice for the laser pulse duration [2]. In particular, this can be achieved by resonant $\pi$-pulse [3]. If the initial atomic momentum $p$ is less than the photon momentum $p_{ph}$, then an atom begins to move in the opposite direction according to the momentum conservation law. Thus, resonant laser light presents a one-dimensional potential barrier for atomic translational motion. The barrier height $V$ is determined by the condition $\sqrt{2MV} = \hbar\omega/c$. For real optical laser frequency $\omega$ and real mass of an atom $M$ this barrier height is much less than 1 K. Therefore an atom should be ultra-cold one, having the kinetic energy on the order of 1 μK. The intense laser light can be also off-resonant. However, then three-dimensional scattering of photons on atoms makes the considered problem more difficult.

Gaseous medium around the considered atom produces the averaged field in Hartree approximation that results in nonlinearity of the Schrödinger single-particle equation for this atom. Tunneling of photon excitations between two Bose-Einstein atomic condensates through the rectangular potential barrier has been investigated in [4]. Various versions of tunneling for the Gross–Pitaevskii equation were considered also in many papers [5–12].

Unlike these works we consider problem of above-barrier reflection of an atom from step potential in the presence of other atoms and derive analytically the reflection coefficient for a weak nonlinearity that corresponds to small number density of atomic gas. This problem was investigated previously in [13–15] using Jacobi elliptic functions.

First we remember the simplest quantum-mechanical problem of above-barrier reflection of a particle with single-particle energy $\mu$ by step potential (see Fig. 2, [16]) for one-dimensional stationary Schrödinger equation (in a system of units where the Planck constant and mass of a particle are equal to unity)

$$-\frac{1}{2}\frac{d^2\psi}{dx^2} + V(x)\psi = \mu\psi;$$

$$V(x) = \begin{cases} 0, & x < 0 \\ V, & x > 0, \end{cases} \quad (1)$$

$$\mu > V.$$

The solution of the stationary Schrödinger equation is the running wave in the direction $x > 0$ which is reflected to the opposite side at $x < 0$

$$\psi(x) = C\exp(ik_2 x); \quad x > 0,$$

$$\psi(x) = \exp(ik_1 x) + b_1\exp(-ik_1 x); \quad x < 0,$$

$$k_1 = \sqrt{2\mu}; \quad k_2 = \sqrt{2(\mu - V)}.$$

---

[1] The article was translated by the author.

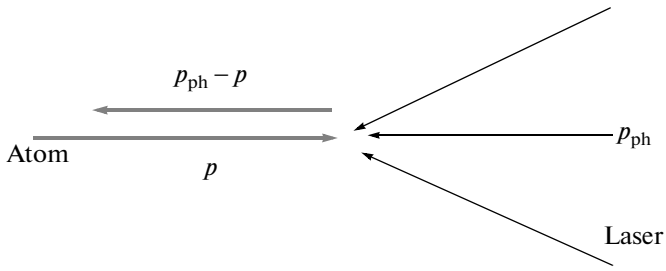

**Fig. 1.** Resonant light as a one-dimensional potential barrier for the translational motion of an atom.

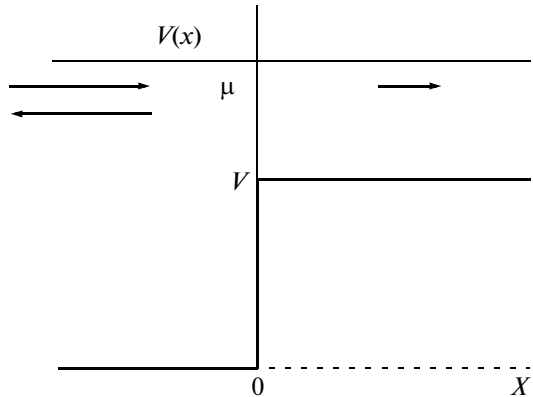

**Fig. 2.** The one-dimensional step potential barrier.

The wave function and its first derivative should be matched at $x = 0$:

$$1 + b_1 = C; \quad k_1(1 - b_1) = k_2 C = k_2(1 + b_1). \quad (2)$$

Hence

$$C = \frac{2k_1}{k_1 + k_2}; \quad b_1 = \frac{k_1 - k_2}{k_1 + k_2} > 0.$$

The reflection coefficient is determined by expression

$$R = |b_1|^2 = \left(\frac{k_1 - k_2}{k_1 + k_2}\right)^2 < 1. \quad (3)$$

Now we consider the Gross–Pitaevskii equation with a weak repulsive ($\alpha > 0$), or attractive ($\alpha < 0$) nonlinearity for the same problem of above-barrier reflection by step potential (attractive nonlinearity is realized for gases from atoms $^7$Li and $^{85}$Rb, while repulsive nonlinearity occurs for a gas from atoms $^{87}$Rb [17]). The temporal Gross–Pitaevskii equation for the single-particle wave function is of the form

$$i\frac{\partial \Psi}{\partial t} = -\frac{1}{2}\frac{d^2 \Psi}{dx^2} + V(x)\Psi + \alpha|\Psi|^2\Psi.$$

The coefficient $\alpha$ is proportional to the amplitude $a_s$ of s-wave elastic scattering of atoms on each other: $\alpha = 4\pi\hbar^2 a_s N/M$ (we assume the short-rang interaction potential and use the Hartree approximation). Here, $N$ is the number of atoms in a gas. The gas is assumed to be sufficiently rarefied, i.e., the scattering length is small in comparison to the distance between atoms. However, this does not mean weak inter-particle interaction since it is proportional to the square of number of atoms, while the kinetic energy is proportional to the number of atoms.

We have for a stationary state with the definite single-particle atomic kinetic energy $\mu$

$$\Psi(x, t) = \psi(x)\exp(-i\mu t).$$

Then the stationary Gross–Pitaevskii equation takes the form

$$-\frac{1}{2}\frac{d^2\psi}{dx^2} + V(x)\psi + \alpha|\psi|^2\psi = \mu\psi. \quad (4)$$

When $x > 0$, a particular solution of Eq. (4) can be written in the same form of the transmitting wave as for the usual Schrödinger equation (it should be noted that since the Gross–Pitaevskii equation is the nonlinear equation, then other solutions can be possible which include the reflecting waves, see [13])

$$\psi(x) = C\exp(ik_2 x); \quad x > 0,$$
$$k_2 = \sqrt{2(\mu - V - |C|^2\alpha)}. \quad (5)$$

The motivation for choice of this solution is based on analogy with a linear case for the transmitting wave.

In the case of $x < 0$ we do not have such simple solution. We solve Eq. (4) at $x < 0$ using the assumption about small dimensionless nonlinear parameter $a = \alpha/\mu \ll 1$. This inequality can be fulfilled at the small atomic number density in a gas. The simple iteration procedure is inapplicable, since so called *secular terms* appear in solution. Namely, non-uniform part of equation contains terms which are simultaneously solutions of the uniform differential equation. These terms result in divergences in iteration procedure. We solve nonlinear Eq. (4) by multiscale analysis which allows us to remove secular terms [18]. This is a form of perturbation theory. The multiscale analysis was applied for the nonlinear Schrödinger equation in [19]. Changing the variable $x \longrightarrow k_1 x$, we rewrite Eq. (4) in the dimensionless form

$$\psi'' + \psi = a|\psi|^2\psi, \quad x < 0. \quad (6)$$

Let us introduce new independent variables

$$x_1 = x, \quad x_2 = ax, \quad x_3 = a^2 x \ldots.$$

Then for the first derivative one obtains

$$\frac{d}{dx} = \frac{\partial}{\partial x_1} + a\frac{\partial}{\partial x_2} + a^2\frac{\partial}{\partial x_3} + \ldots,$$

while for the second derivative we have

$$\frac{d^2}{dx^2} = \frac{\partial^2}{\partial x_1^2} + 2a\frac{\partial^2}{\partial x_1 \partial x_2} + \ldots. \quad (7)$$

Here and thereafter we restrict ourselves only by terms which are linear with respect to the small dimensionless parameter $a \ll 1$.

Substituting Eq. (7) into Eq. (6), we find the nonlinear equation

$$\frac{\partial^2 \psi}{\partial x_1^2} + 2a\frac{\partial^2 \psi}{\partial x_1 \partial x_2} + \psi = a|\psi|^2\psi. \quad (8)$$

According to general rules of the iteration procedure, the solution of Eq. (8) should be expanded on the small parameter $a$

$$\psi = \psi_0 + a\psi_1 + a^2\psi_2 + \ldots.$$

Substituting this solution into Eq. (8), one obtains first the equation for the zero iteration term

$$\frac{\partial^2 \psi_0}{\partial x_1^2} + \psi_0 = 0.$$

Its solution is of a simple form

$$\psi_0(x_1, x_2) = A(x_2)\exp(ix_1) + B(x_2)\exp(-ix_1). \quad (9)$$

Now we consider equation for the first iteration term $\psi_1$. It follows from Eq. (8) that

$$\frac{\partial^2 \psi_1}{\partial x_1^2} + 2\frac{\partial^2 \psi_0}{\partial x_1 \partial x_2} + \psi_1 = |\psi_0|^2\psi_0. \quad (10)$$

Substituting Eq. (9) for zero iteration term into Eq. (10), one obtains the non-homogeneous linear differential equation for the function $\psi_1$ (primes mean derivatives)

$$\frac{\partial^2 \psi_1}{\partial x_1^2} + \psi_1 = |\psi_0|^2\psi_0 - 2\frac{\partial^2 \psi_0}{\partial x_1 \partial x_2}$$
$$= [|A|^2 + |B|^2 + AB^*\exp(2ix_1) + A^*B\exp(-2ix_1)] \quad (11)$$
$$\times (A(x_2)\exp(ix_1) + B(x_2)\exp(-ix_1))$$
$$- 2i(A'(x_2)\exp(ix_1) - B'(x_2)\exp(-ix_1)).$$

It is seen that the non-homogeneous part which is proportional to the exponent $\exp(ix_1)$, is the solution of the uniform differential equation for the function $\psi_1$. This part produces secular term which increases with the increase of $x_1$ up to infinity. Analogous secular term is produced by the non-homogeneous part which is proportional to the exponent $\exp(-ix_1)$. We should equalize these two terms to zero. Then according to Eq. (11) next conditions should be fulfilled:

$$A[|A|^2 + 2|B|^2] = 2iA';$$
$$B[2|A|^2 + |B|^2] = -2iB'. \quad (12)$$

In order to solve these ordinary differential equations, we present the complex quantity $A$ and $B$ in standard form of complex numbers

$$A = a_1\exp(ia_2);$$
$$B = b_1\exp(ib_2).$$

Then the system of Eqs. (12) can be immediately solved:

$$a_1' = 0; \quad b_1' = 0; \quad a_1 = \text{const}; \quad b_1 = \text{const};$$
$$a_1^2 + 2b_1^2 = -2a_2'; \quad a_2 = -(a_1^2/2 + b_1^2)x_2;$$
$$2a_1^2 + b_1^2 = 2b_2'; \quad b_2 = (a_1^2 + b_1^2/2)x_2.$$

Thus, according to Eq. (9) one obtains zero iteration for wave function (we return to the previous variables)

$$\psi_0(x) = a_1\exp(iv_1x) + b_1\exp(-iv_2x);$$
$$v_1 = k_1[1 - a(a_1^2/2 + b_1^2)]; \quad (13)$$
$$v_2 = k_1[1 - a(a_1^2 + b_1^2/2)].$$

We put the constant $a_1 = 1$ without restriction of general form of solutions analogously to such operation in the usual Schrödinger equation for the problem of transmission and reflection of plane waves. Then the quantity $b_1$ can be defined as the reflection amplitude, and its square is the reflection coefficient.

Now in the right side of Eq. (11) only third harmonics remain:

$$\frac{\partial^2 \psi_1}{\partial x_1^2} + \psi_1 = A^2B^*\exp(3ix_1) + A^*B^2\exp(-3ix_1).$$

In the right side of this expression we should substitute zero approximations for the quantities $A$ and $B$, i.e., $A = 1$, $B = b_1$. Hence, Eq. (11) takes the form

$$\frac{\partial^2 \psi_1}{\partial x_1^2} + \psi_1 = b_1\exp(3ix_1) + b_1^2\exp(-3ix_1).$$

The partial solution of this equation is of a form

$$\psi_1 = -\frac{1}{8}b_1\exp(3ix_1) - \frac{1}{8}b_1^2\exp(-3ix_1).$$

Thus, taking first iterations into account, one finds the whole solution

$$\psi(x) = \exp(iv_1x) + b_1\exp(-iv_2x)$$
$$- \frac{a}{8}b_1\exp(3ik_1x) - \frac{a}{8}b_1^2\exp(-3ik_1x).$$

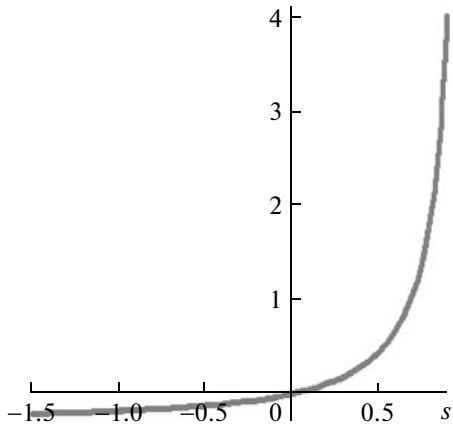

**Fig. 3.** Plot of the function $F(s)$.

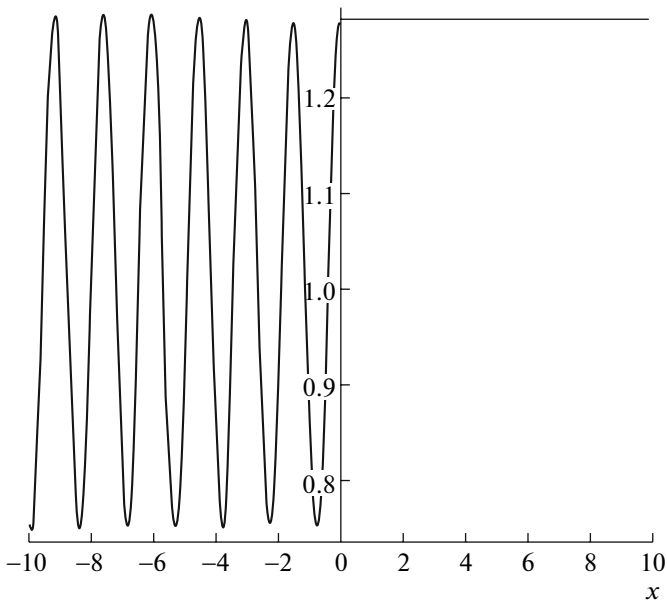

**Fig. 4.** Stationary solution of the Schrödinger equation for the probability density $\rho(x) = |\psi(x)|^2$.

Now we match the wave function and its first derivative at $x = 0$. The one obtains equations for determination of constants $C$ and $b_1$

$$1 + b_1(1 - a/8 - ab_1/8) = C;$$

$$1 - a(1/2 + b_1^2) - b_1[1 - a(1 + b_1^2/2)]$$
$$- 3ab_1/8 + 3ab_1^2/8 = k_2 C/k_1.$$

Excluding the constant $C$, we find equation for determination of the reflection amplitude $b_1$

$$1 - a(1/2 + 5b_1^2/8) - b_1[1 - a(5/8 + b_1^2/2)]$$
$$= k_2[1 + b_1(1 - a/8 - ab_1/8)]/k_1.$$

When $a = 0$ this equation reduces to Eq. (2) as it should be. Further we rewrite this equation in the form

$$1 - b_1 - k_2(1 + b_1)/k_1 = a(1/2 + 5b_1^2/8) \quad (14)$$
$$- ab_1(5/8 + b_1^2/2) - ak_2 b_1(1 + b_1)/8k_1.$$

In the right side of this equation which is proportional to the small nonlinear parameter $a$, we should substitute the value of $b_1$, which does not take into account the nonlinearity, i.e.,

$$b_1 = \frac{k_1 - k_2}{k_1 + k_2} = \frac{1 - \sqrt{1-s}}{1 + \sqrt{1-s}}; \quad (15)$$
$$s = V/\mu.$$

In the left side of Eq. (14) we should substitute

$$k_2/k_1 = \sqrt{1 - s - a|C|^2} \approx \sqrt{1-s} - a\frac{(1+b_1)^2}{2\sqrt{1-s}}.$$

Taking into account this relation, we rewrite Eq. (14) in the form

$$1 - b_1 - \sqrt{1-s}(1 + b_1) = a(1/2 + 5b_1^2/8)$$
$$- ab_1(5/8 + b_1^2/2) - ak_2 b_1(1+b_1)/8k_1 - a\frac{(1+b_1)^3}{2\sqrt{1-s}}.$$

Hence,

$$b_1 = \frac{1 - \sqrt{1-s}}{1 + \sqrt{1-s}} + aF(s).$$

Here the function $F(s)$ is determined

$$F(s)(1 + \sqrt{1-s}) = -1/2 + 5b_1/8 - 5b_1^2/8 + b_1^3/2$$
$$+ b_1(1+b_1)\sqrt{1-s}/8 + \frac{(1+b_1)^3}{2\sqrt{1-s}}. \quad (16)$$

Substituting Eq. (15) into Eq. (16), we simplify the expression for the function $F(s)$

$$F(s) = \frac{s(13 - 5s)}{2\sqrt{1-s}(1+\sqrt{1-s})^4} > 0 \quad \text{at} \quad x > 0. \quad (17)$$

Thus, the repulsive nonlinearity ($a > 0$) increases reflection from potential barrier while the attractive nonlinearity ($a < 0$) decreases reflection and increases transmission through step potential [13]. Besides of this, incident and reflecting waves have different effective masses because of the nonlinearity. We determine here incident and reflecting waves as parts of the wave function which reduce to $\exp(ik_1 x)$ and $\exp(-ik_1 x)$, respectively, in linear case ($\alpha = 0$).

Plot of the function $F(s)$ is presented in Fig. 3. It is seen that when the kinetic energy of an atom increases compared to the height of the potential barrier $V$ (i.e., when $s$ decreases), the role of nonlinearity diminishes.

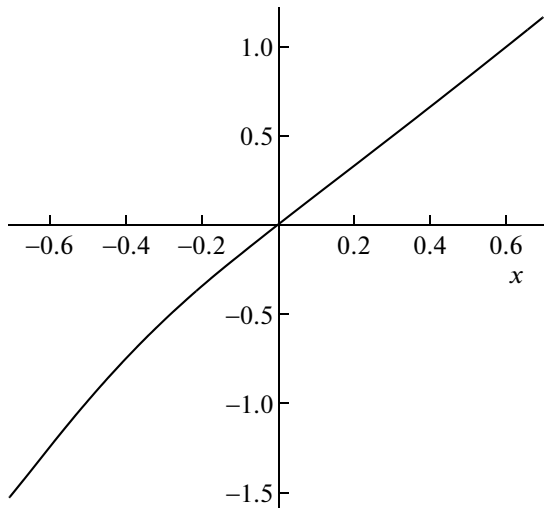

**Fig. 5.** Phase of the wave function $\theta(x)$.

The solution (17) is valid also for the case when $V < 0$, that corresponds to the reflection by potential well. From the physical point of view this corresponds to a laser photon which initially moves in the same direction as an atom (i.e., in positive direction of the axis $X$). The role of nonlinearity decreases in this case as it should be.

If $\mu = V$, then according to (5) transmission through barrier takes place only for attractive nonlinearity ($a < 0$). In accordance with Eq. (14) we obtain amplitude of reflection

$$b_1 \approx 1 - 4\sqrt{|a|}. \qquad (18)$$

In this case small transmission coefficient is equal to

$$T = 1 - b_1^2 \approx 8\sqrt{|a|}.$$

For attractive nonlinearity transmission through step potential is possible also at $\mu < V$ (see also [13]).

Oppositely, for repulsive nonlinearity according to Eq. (5) transmission through barrier begins not when $\mu = V$, but for the definite energy $\mu_0 > V$. This value of $\mu_0$ can be derived from relation (again when $a \ll 1$)

$$1 - V/\mu_0 = a|C|^2 \approx a\frac{4}{(1 + \sqrt{1 - V/\mu_0})^2} \approx 4a \ll 1; \qquad (19)$$
$$\mu_0 - V \approx 4\alpha.$$

In Fig. 4 we present the probability density $\rho(x) = |\psi(x)|^2$ for the example $V = 1$, $\mu = 2.4$, $\alpha = 0.2$, which was considered numerically in [13]. This quantity is given by simple expression based on the above approach:

$$\rho(x) = 1.026 + 0.320\cos(4.094x) - 0.002\cos(4.477x),$$
$$x < 0; \qquad (20)$$
$$\rho(x) = 1.344, \quad x > 0.$$

The phase of the wave function $\theta$, which is determined from the dependence $\psi(x) = \sqrt{\rho(x)}\exp(i\theta)$, is shown in Fig. 5. The phase is given by the expression

$$\theta(x) = \arctan[0.764\tan(2.191x)], \quad x < 0;$$
$$\theta(x) = 1.673x, \quad x > 0. \qquad (21)$$

The last term in the right side of the expression for $\rho(x)$ in Eq. (20) corresponds to contribution of the third harmonics. It is seen that it is very small. Therefore the incident and reflecting waves are described well by the function (13) when $x < 0$.

In Fig. 6 we present the numerical solution for the probability density $\rho(x)$ from [13] derived for the same values of parameters (but in other system of units both for $x$ and for density). The difference between Fig. 4 and Fig. 6 is explained by different boundary conditions in our approach and in [13]. We started with the linear problem of above-barrier reflection when on the region $x > 0$ there is only the transmitting wave. We assumed that reflecting wave does not appear in the nonlinear problem at $x > 0$. It is seen from Fig. 6 of [13] that the transmitting wave is modulated by small reflection produced only by the nonlinearity. This results in rare periodic strong decreasing of the density at large values of $x > 0$ (Fig. 6).

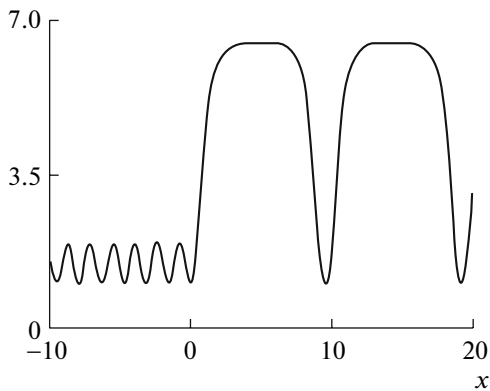

**Fig. 6.** The probability density $\rho(x) = |\psi(x)|^2$ according to [13].

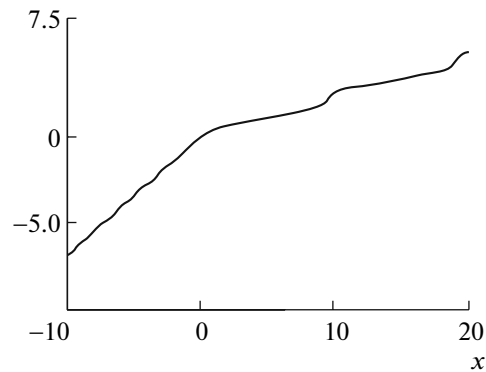

**Fig. 7.** Phase of the wave function $\theta(x)$ according to derivations of [13].

The numerical dependence of the phase $\theta$ on $x$ is presented in Fig. 7 according to derivations of [13].

Generally, it is reproduced qualitatively in our approach (see Fig. 5). It should be noted that qualitative agreement of analytical and numerical solutions is explained by small value of the nonlinear parameter $\alpha = 0.2$ in the Gross–Pitaevskii equation for the considered example.

This work was supported by RFBR (grant no. 07-02-00080).